\documentclass[twocolumn,resetfootnote]{aastex701}
\let\tablenum\relax
\usepackage{siunitx}
\usepackage[version=4]{mhchem} 
\usepackage{amsmath}
\usepackage{amssymb}

\usepackage{cleveref}

\DeclareSIUnit{\Gyr}{Gyr}
\DeclareSIUnit{\erg}{erg}

\begin{document}

\title{Coupled Thermal--Chemical Evolution Models of Sub-Neptunes Reveal Atmospheric Signatures of Their Formation Location}

\author[orcid=0000-0003-0605-0263,sname='Steinmeyer',gname='Marie-Luise']{Marie-Luise Steinmeyer}
\affiliation{Institute for Particle Physics and Astrophysics, ETH Zurich, CH-8093 Zurich, Switzerland}
%\correspondingauthor{Marie-Luise Steinmeyer}
\email[show]{steinmeyer\_ml@yahoo.com}  

\author[orcid=0000-0001-6110-4610,sname='Dorn',gname='Caroline']{Caroline Dorn} 
\affiliation{Institute for Particle Physics and Astrophysics, ETH Zurich, CH-8093 Zurich, Switzerland}
\email{dornc@ethz.ch}

\author[orcid=0009-0005-1133-7586,sname='Werlen',gname='Aaron']{Aaron Werlen} 
\affiliation{Institute for Particle Physics and Astrophysics, ETH Zurich, CH-8093 Zurich, Switzerland}
\email{aaron.werlen@icloud.com}

\author[orcid=0000-0002-0632-4407,sname='Grimm',gname='Simon L.']{Simon L. Grimm}
\affiliation{Institute for Particle Physics and Astrophysics, ETH Zurich, CH-8093 Zurich, Switzerland}
\affiliation{Department of Astrophysics, University of Zurich, CH-8057 Zurich, Switzerland}
\email{sigrimm@ethz.ch}

% \collaboration{20}{(AAS Journals Data Editors)}



\begin{abstract}
The observed masses and radii of sub-Neptunes are typically explained by the gas dwarf or water world scenarios.
While their evolutionary history on a population level has been proposed as a method to distinguish between these compositions, previous evolutionary models neglected the crucial role of atmosphere–interior chemical interaction.
We present a novel evolution framework for sub-Neptunes that combines the thermal evolution with the chemical coupling of the atmosphere and interior. 
Using this model, we examine how planets formed
inside and outside the water-ice line can be observationally distinguished, with an emphasis on their atmospheric properties.
We find that young sub-Neptunes store the majority of their volatile budget in the interior, regardless of formation location. 
Nevertheless, the atmospheric metallicity is a factor of 4 higher for the planet formed outside the water-ice line.
During cooling, hydrogen and oxygen exsolve from the interior, increasing the
atmospheric mass fraction and counteracting the thermal contraction for both planets. 
Consequently, radius evolution alone cannot distinguish between the two formation scenarios.
Instead, the primary discriminators are the abundance of carbon-bearing species and the resulting atmospheric C/O ratio.
For sub-Neptunes formed beyond the water-ice line, nearly all carbon resides in the gaseous phase. We find that high molar fractions of \ce{CH4} ($>10^{-2}$) and \ce{H2O} ($> 5\times10^{-2}$), and a high C/O ratio $(> 5\times10^{-1})$ are indicative of formation outside the water-ice line.
In contrast, sub-Neptunes formed inside the water-ice line exhibit strongly suppressed \ce{CH4} abundances, yielding C/O ratios ranging from $10^{-7}$ to $10^{-1}$.
\end{abstract}


%% We recommend that authors also use the natbib \citep and \citet commands to identify citations. The citations are
%% tied to the reference list via symbolic KEYs. The KEY corresponds to the KEY in the \bibitem in the reference list below. 

\section{Introduction}
Around $50\%$ of Sun-like stars are expected to be orbited by at least one sub-Neptune with a radius between $\sim 2$ and $4\,R_\oplus$ and orbital periods $\lesssim 100\,$days \citep{2013_Fressin_occurence,2013_Petigura_prevalence,2013_Batalha_Kepler,2019_He_Kepler}.
Their precise internal structure and composition remain ambiguous, but their low bulk density requires that they contain a thick volatile layer, giving rise to two leading composition scenarios: the gas dwarf scenario and the water world scenario.
In the first scenario, sub-Neptunes are planets with a rocky interior that accreted thick \ce{H2}/He-dominated atmospheres\footnote{Here, we use the terms atmosphere and envelope interchangeably to refer to everything above the silicate layer}. \citep[e.g.,][]{2014Lee,ginzburg_super-earth_2016}.
In the second scenario, sub-Neptunes are extremely water-rich, with water mass fractions up to $50\,\%$ of the total planet mass \citep{zeng_growth_2019,2020Mousis_Irradiatedoceans,luque_density_2022}.
Recent atmospheric characterizations of sub-Neptune atmospheres by JWST reveal a large diversity in composition, from \ce{H2}/He-dominated atmospheres with solar metallicity \citep{2025DavenportTOI} to high-metallicity or even water-dominated atmospheres \citep[e.g.,][]{2023Kempton,2024Beattysulfur,benneke_jwst_2024,piaulet-ghorayeb_jwstniriss_2024}. 
This highlights that both the gas dwarf and water world scenarios are likely extreme end-member cases. 
Sub-Neptunes likely accrete both \ce{H2}/He and \ce{H2O}, especially when they preferentially form outside the water-ice line \citep{venturini_nature_2020,burn2024water, chakrabarty2024water}.

Both composition scenarios can explain the observed masses and radii of sub-Neptunes, making it difficult to distinguish between them based on bulk density alone \citep[e.g.,][]{2011_Rorgers_exoNeptunes,dorn_can_2015,2015_Rogers_most,bean_nature_2021}. 
To break this degeneracy, recent works have proposed to examine the evolution of sub-Neptunes over time \citep{aguichine_evolution_2024,rogers_road_2025}. 
Sub-Neptunes of different compositions are expected to cool and contract differently over time.  %
In particular, gas-rich sub-Neptunes experience substantial radius shrinkage as they cool \citep{nettelmann_thermal_2011,misener_cool_2021,rogers_road_2025}, whereas water-rich planets exhibit only modest changes in radius over time  \citep{aguichine_evolution_2024}. 
Consequently, it may be possible to distinguish between the two composition scenarios statistically across the growing population of sub-Neptunes spanning a wide range of ages, especially with the expanding TESS sample \citep{rogers_road_2025}.

However, these models unrealistically assume a complete decoupling of the atmosphere from the interior, with all volatiles confined to the envelope. 
This simplification disregards the potential for chemical exchange between the atmosphere and the deep interior, which has been proposed to crucially alter the composition of the interior and the atmosphere of a planet \citep{schlichting2022,2024_Seo_subNeptune, 2025HengGradient,werlen_sub-neptunes_2025,werlen_atmospheric_2025}. 
The deep interiors of sub-Neptunes are expected to host magma oceans that are vigorously convecting and well mixed \citep[e.g.,][]{2021Lichtenbergredox}, enabling efficient interaction with the overlying gas layer.
Because the thick atmospheres of sub-Neptunes provide strong thermal insulation, these magma oceans may persist for extended timescales, allowing continuous atmosphere–interior chemical exchange \citep[e.g.,][]{2016Schaefer_predictions,2019KiteSuperabundance,2020KiteAtmosphere,2024Nicholss_magma,2025_Tang_sub-Neptune}. 

Most importantly, this chemical coupling leads to the partitioning of volatiles into the deep interior of the planet, which reduces the atmosphere mass and total radius of the planet \citep{dorn_hidden_2021,schlichting2022,young_earth_2023,luo_majority_2024,2025_Bower_Diversity}. 
The resulting atmosphere mass is strongly dependent on the temperature at the atmosphere-magma ocean interface (AMOI) \citep{schlichting2022}.
The chemical coupling further leads to an increase in heavier gas species. 
Chemical reactions between hydrogen and magma ocean lead to endogenic water production \citep{2021_Kite_rocky,schlichting2022,kimura_predicted_2022, young_earth_2023}. 
At the same time, rock vapor is stable in the atmosphere due to the high temperature at the AMOI \citep{2013VisscherChemistry,2022Misener}.
Combined, these effects fundamentally alter the composition and mass of the atmosphere, which influences how efficiently a planet can cool. 
The temperature at the AMOI, in itself, is dictated by the planet's evolution. 
However, the interplay of planetary evolution and chemical coupling has not yet been studied in detail. 

A key step in this direction was recently taken by \citet{rogers_redefining_2025}, who present an evolutionary framework for sub-Neptunes that takes into account the miscibility of \ce{MgSiO3} and \ce{H2}. 
They find a significant reduction of the atmosphere mass fraction for young sub-Neptunes, although the majority of \ce{H2} subsequently exsolves as the planet cools. 
However, \citet{rogers_redefining_2025} only consider pure \ce{MgSiO3}-\ce{H2} planets and thus neglect the effects of additional species such as Fe or \ce{H2O}. 

In this work, we present a novel evolution model that incorporates global chemical equilibrium calculations within a thermal evolution framework for sub-Neptunes. 
This allows us to track the distribution of elements across the metallic, silicate, and gaseous phase of the planet over time. 
Recent works suggest that sub-Neptunes may not differentiate into a discrete silicate mantle and iron core, as is the case for the terrestrial planets in the solar system, but that the metallic and silicate phases remain mixed throughout the planet's evolution \citep{wahl2015high,insixiengmay2025mgo,2025young_differentiation}
Therefore, we consider the extreme end-member case where the metallic phase can participate in the chemical exchange throughout the entire evolution. 
We further neglect atmospheric escape and miscibility between silicates, iron, and hydrogen in order to isolate the effect of the chemical coupling between the atmosphere and the interior on the evolution.

We compare the evolution pathways of two distinct types of sub-Neptunes: planets formed inside the water-ice line that accrete only \ce{H2}/He and planets formed outside the water-ice line that accrete both \ce{H2}/He and \ce{H2O}.
We explore the differences in their cooling and contraction histories, as well as in the chemical composition of their atmospheres. 
Ultimately, our goal is to identify observable properties that can distinguish their initial bulk compositions and thereby their formation locations.

This paper is structured as follows: In Section \ref{sec:method}, we describe the underlying planetary structure model, the thermal evolution framework, and the global chemical equilibrium model. 
In Section \ref{sec:results}, we show the impact of coupled atmosphere--interior models on the evolution of sub-Neptunes. We validate our results considering a larger compositional diversity and discuss the implications for observations in Section \ref{sec:Discussion}. 
Finally, the conclusions are provided in Section \ref{sec:conclusions}.

\section{Method}
\label{sec:method}
The coupled atmosphere--interior model consists of three separate elements. 
The first is the calculation of the structure of the planet (Section \ref{ssec:structure}), followed by the thermal evolution (Section \ref{ssec:thevolution}), and the calculation of the global chemical equilibrium of the planet (Section \ref{ssec:gce}). 
We discuss the general workflow of how the framework combines these three elements in Section \ref{ssec:workflow}

\subsection{Planetary Structure Model}
\label{ssec:structure}
We follow and adapt the interior model of \citep{dorn_generalized_2017,dorn_hidden_2021}.
We assume a planet that is spherically symmetric and in hydrostatic equilibrium. 
The structure of a planet is then calculated by solving the following equations
\begin{align}
    \frac{dP}{dm} &= - \frac{Gm}{4\pi r^4} \label{eq:presstruceq} \\
    \frac{dr}{dm} &=  \frac{1}{4\pi r^2 \rho} \\
    \frac{dT}{dr} &= \frac{T}{P} \frac{dP}{dr} \nabla(T,P).
    \label{eq:structureeq}
\end{align}
Here, $P$ is the pressure, $m$ is the mass inside the radius $r$ from the planet's center, $\rho$ is the density at radius $r$, $T$ the temperature, and $\nabla(T,P) = d \ln T / d \ln P$. 
We divide the planet from the center outward into an iron core, a silicate mantle, and an atmosphere in radiative-convective equilibrium. 
In the following, we will use the term \textit{interior} to refer to the metallic and silicate phases of the planet combined. 

The atmosphere is further divided into the upper atmosphere, where most of the stellar irradiation is absorbed and the lower atmosphere, which is only heated by the thermal energy from the interior. 
If the upper atmosphere is convectively stable, we use the temperature profile for the global average by \citet{guillot_radiative_2010}, which is a semigray model using the Eddington approximation that takes horizontal advection into account
\begin{equation}
    \begin{split}
        T^4 =& \frac{3 T_\mathrm{int}^4}{4}\Bigl\{\frac{3}{2} + \tau \Bigr\} +\frac{3 T_\mathrm{eq}^4}{4}\Bigl\{\frac{3}{2} + \frac{2}{3\gamma} \left[1 + \left(\frac{\gamma \tau }{2}-1\right)e^{\gamma \tau}\right] \\
        &+ \frac{2\gamma }{3} \left(1-\frac{\tau^2}{2}\right) E_2(\gamma \tau)\Bigr\}.
    \end{split}
\end{equation}
Here, the heat flux from the planet's interior is given by the intrinsic temperature $T_\mathrm{int} = (L_\mathrm{int}/4\pi\sigma_B R_\mathrm{pl}^2)^{1/4}$, with $L_\mathrm{int}$ the intrinsic luminosity of the planet, and $R_\mathrm{pl}$ the total planet radius. 
The equilibrium temperature $T_\mathrm{eq} = T_* (R_*/2D)^{1/2}$ depends on the stellar effective temperature $T_*$, stellar radius $R_*$, and the orbital distance $D$ of the planet. 
The opacity is incorporated by the ratio of the visible and thermal opacity $\gamma = \kappa_\mathrm{v}/\kappa_\mathrm{th}$ and $\tau$ is the optical depth. 
This parameter determines how much flux is absorbed in the upper atmosphere. 
We use the tabulated values from \citet{jin_planetary_2014} to evaluate $\gamma$. 
$E_2(\gamma \tau$) is the second-order exponential integral $E_2(z) = \int_1^\infty t^{-1}e^{-zt} dt$. 
The gradient of the optical depth is given by 
\begin{equation}
    \frac{\mathrm{d} \tau}{\mathrm{d} r} = \kappa_\mathrm{th}\rho.
\end{equation}
The temperature at the upper boundary of the atmosphere, where $P_\mathrm{TOA}=10^{-10}\,$bars, is set by 
\begin{equation}
    T_\mathrm{TOA} = \frac{3}{4} \left(\frac{2 \sqrt{2}T_\mathrm{eq}^4}{3} + \frac{T_\mathrm{irr}}{4} \left(\frac{2}{3}+\frac{\gamma}{\sqrt{3}}\right)\right).
\end{equation}
The transit radius of the planet is defined as the radius at which the chord optical depth is $\tau_\text{ch}=0.56$ \citep{2008Lecavelier}. 

The transition between the irradiated, upper atmosphere  and the lower nonirradiated atmosphere happens when essentially all stellar flux is absorbed by the atmosphere above, i.e., when $\tau_v \gg 1$. 
We set the boundary at $\tau = 100/(\sqrt{3}\gamma)$ following \citet{jin_planetary_2014}. 
If a layer in the lower atmosphere is convectively stable, we use the radiative temperature gradient
\begin{equation}
\nabla_\text{rad} = \frac{3}{16 \pi a c G}\frac{\kappa_\mathrm{th} L_\text{int} P}{m T^4}, \label{Eq:rad_grad}
\end{equation}
where $a$ is the radiation density constant, $c$ the speed of light, and $G$ the gravitational constant.
For the thermal opacity, we use $\kappa_\mathrm{th} = \max(0.01 \si{\centi\meter\squared\per\gram},\kappa_\mathrm{F})$, where $\kappa_\mathrm{F}$ is the Rosseland mean opacities by \citet{freedman_gaseous_2014}.
The minimum value in the thermal opacity is set to match the constant thermal opacity used in the semigray model of the irradiated upper atmosphere by \citet{guillot_radiative_2010}. 

For convectively unstable layers, the temperature gradient is given by the adiabatic gradient
\begin{equation}
\nabla_\text{ad} = \left(\frac{d \ln(T)}{d \ln(P)}\right)_S. 
\end{equation}
The adiabatic temperature gradient and other thermodynamic quantities are calculated by mixing the equation of state (EOS) for a H/He-mixture by \citet{1995_Saumon_EOS} and the ANEOS EOS for water \citep{1990_thompson_aneos} using the additive volume law.

The conditions at the bottom of the atmosphere are used as the outer boundary for the interior of the planet. 
We assume an adiabatic temperature profile and consider both liquid and solid phases for the core and mantle. 
For liquid iron, we use the EOS by  \citet{luo_majority_2024}, while for solid iron we use the EOS for hexagonal close-packed iron \citep{hakim_new_2018,miozzi_new_2020}. 
The solid mantle mineralogy is based on the three major species MgO, \ce{SiO2}, and FeO. 
For pressures below $\approx 125\,$GPa, we use the thermodynamical model \textsc{Perple\_X} \citep[e.g.,][]{connolly_geodynamic_2009} to calculate the exact mineralogy, while for higher pressures, we define the stable minerals a priori and use their respective EOS \citep{hemley_constraints_1992,fischer_equation_2011,faik_equation_2018,musella_physical_2019,luo2023equation}.
The liquid mantle is modeled as a mixture of \ce{Mg2SiO4}, \ce{SiO2} and FeO since there are no data available for the density of liquid MgO in the required pressure-temperature regime \citep{melosh_hydrocode_2007,faik_equation_2018,ichikawa_ab_2020,stewart_shock_2020}.
All components are mixed using the additive volume law. 

\subsection{Thermal Evolution}
\label{ssec:thevolution}
We model the thermal evolution using the total energy conservation approximation adopted in previous works \citep[e.g.,][]{mordasini_characterization_I_2012,rogers_road_2025}, 
\begin{equation}
    \frac{d E_\mathrm{tot}}{d t} = -L,
    \label{eq:dEdt}
\end{equation}
where $E_\mathrm{tot}$ is the total energy of the planet and $L$ the total luminosity of the planet. 

The total energy of the planet is the sum of the gravitational and internal (thermal) energy of the planet interior and atmosphere: 
\begin{equation}
    E_\mathrm{tot} = E_\mathrm{grav,int} + E_\mathrm{th,int} + E_\mathrm{grav,atm} + E_\mathrm{th,atm}
    \label{eq:Etot}
\end{equation}
The total energy is calculated following the approach presented in \citet{linder_evolutionary_2019}. 
In order to decrease computation time, the energy contribution of the rocky interior is modeled assuming an isothermal sphere of constant density.
More specifically, the thermal energy of the rocky interior is 
\begin{equation}
    E_\mathrm{th,int} = c_{v,\mathrm{int}}M_\mathrm{int}T_\mathrm{AMOI}
    \label{eq:ethint}
\end{equation}
with $M_\mathrm{int}$ the mass of the interior and $T_\mathrm{AMOI}$ the temperature at the atmosphere--interior interface. 
For the heat capacity of rocky material, we use $c_{v,\mathrm{int}}=10^7\,\si{\erg\per\gram\per\K}$ \citep{1995Guilloteffect,linder_evolutionary_2019}. 
Assuming an isothermal temperature profile underestimates the contribution of the interior to the total energy compared to a fully convective interior; however, \citet{baraffe_structure_2008} and \citet{linder_evolutionary_2019} showed that its impact on the evolution of the planet is negligible.  
Under the assumption of a constant density for the interior, the gravitational energy of the planet's interior is calculated as
\begin{equation}
    E_\mathrm{grav,int} = - \frac{3\mathrm{G}M_\mathrm{int}^2}{5 R_\mathrm{int}}.
\end{equation} 

The thermal energy of the atmosphere is given by
\begin{equation}
    E_\mathrm{th,atm} = \int_{M_\mathrm{int}}^{M_\mathrm{pl}} u_\mathrm{atm} dm,  
\end{equation}
where $u$ is the specific thermal energy taken from the mixed EOS for H-He-\ce{H2O}. 
Lastly, the gravitational energy of the atmosphere is
\begin{equation}
    E_\mathrm{grav,atm} =  - \int_{M_\mathrm{int}}^{M_\mathrm{pl}}  \frac{G m(r)}{r} dm
\end{equation}

Next to the cooling and contraction of the atmosphere, we include the heating by the decay of radioactive material in the silicate mantle as an additional source of luminosity. 
The radioactive luminosity is given by
\begin{equation}
    L_\mathrm{radio}(t) = Q_\mathrm{tot}(t)f_\mathrm{mantle}M_\mathrm{int},
\end{equation}
where $Q_\mathrm{tot}(t)$ is the heating rate per gram, $f_\mathrm{mantle}$ is the mantle mass fraction, and $M_\mathrm{int}$ the mass of the interior. 
We follow \citet{mordasini_characterization_II_2012} and calculate the heating rate assuming a chondritic abundance of radionucleids,
\begin{equation}
    Q_\mathrm{tot}(t) = Q_{0,\mathrm{K}}\mathrm{e}^{-\lambda_\mathrm{K}t} + Q_{0,\mathrm{U}}\mathrm{e}^{-\lambda_\mathrm{U}t} + Q_{0,\mathrm{Th}}\mathrm{e}^{-\lambda_\mathrm{Th}t}.
\end{equation}
Here, $Q_{0,i}$ is the heating rate at $t=0$ and $\lambda_i$ the decay constant for each nuclide $i$. 
The values for $Q_{0,i}$ and $\lambda_i$ are taken from \citet{mordasini_characterization_II_2012}, who derived the values from the data given in \citet{lowrie2007fundamentals}. They are listed in Table \ref{tab:radiodata}.
The selection of radionuclids is based on \citet{1964Wasserburg}.
\begin{deluxetable}{lcc}
\tablecaption{Parameters for radioactive heating}
\label{tab:radiodata}
\tablehead{\colhead{Nuclide} & \colhead{$Q_0$ [erg/(g s)]} & \colhead{$\lambda$ [1/Gyr]}}
\startdata
\ce{^40K} &  $3.723\times 10^{-7}$ & 0.543\\
\ce{^238U} & $2.899\times 10^{-8}$ & 0.155 \\
\ce{^232Th} & $1.441 \times 10^{-8}$ & 0.0495
\enddata
\tablecomments{Values taken from \citet{mordasini_characterization_II_2012} based on data from \citet{lowrie2007fundamentals}.}
\end{deluxetable}

We benchmark our thermal evolution model by comparing it to the evolutionary models of \citet{vazan_contribution_2018} in the appendix \ref{app:benchmark}.

\subsection{Global Chemical Equilibrium}
\label{ssec:gce}
We employ the global chemical equilibrium framework (GCE) described in \citet{werlen_atmospheric_2025, werlen_sub-neptunes_2025}, which builds on the formulation of \citet{schlichting2022} and \citet{young_earth_2023} and extend it by including carbon partitioning into the metallic phase. The framework distributes 26 components across the metal, silicate, and gas phases by solving a network of 19 independent reactions. For each reaction, equilibrium is imposed by the condition  

\begin{equation}\label{eq:chemical_equilibrium}
    \sum_i \nu_i \ln x_i + \left[\frac{\Delta \hat{G}^\circ_{\text{rxn}}}{RT} + \sum_g \nu_g \ln(P/P^\circ)\right] = 0,
\end{equation}

\noindent where $x_i$ is the mole fraction of species $i$ in its respective phase and $\nu_i$ the associated stoichiometric coefficients. The term $\Delta \hat{G}^\circ_\text{rxn}$ is the standard Gibbs free energy change of the reaction, $R$ the universal gas constant, $T$ the temperature, $P$ the pressure at the AMOI, and $P^\circ$ the reference pressure (set to 1~bars). Pressure corrections are included only for gaseous species $g$, introducing the explicit pressure dependence in the equilibrium relation. The full chemical reaction network is provided in the Appendix of \citet{werlen_sub-neptunes_2025}.

The thermodynamic data used to calculate Gibbs free energies are taken from the NIST database, the MAGMA code \citep{fegley_vaporization_1987} where available, or from the literature. A detailed description is provided in the appendix of \citet{schlichting2022}. In addition, we include activity coefficients to account for the non-ideal behavior of \ce{Si} and \ce{O} in metal, following \citet{young_earth_2023}, \citet{werlen_sub-neptunes_2025} and \citet{werlen_effects_2026} and based on the expressions of \citet{badro_core_2015}.
The calculation of the Gibbs free energy of \ce{H2} dissolved in silicates is updated to include the data from \citet{gilmore_core-envelope_2026}.
Furthermore, we now account for the pressure dependence of the partitioning of hydrogen between metal and silicate. 
As a result, the mass fraction of \ce{H2} in the silicate phase is increased compared to previous versions of the GEC, in line with recent experimental results by \citet{2025Miozzi}.
For more details on the treatment of hydrogen in the silicate and metal phase see \citet{werlen_effects_2026}.

The system of equations is solved following the numerical approach introduced by \citet{schlichting2022}, but with substantial performance optimizations described in S. L. Grimm et al. (in prep.).

\subsection{Planet Sample and Simulation Workflow}
\label{ssec:workflow}
\begin{deluxetable}{lcccccc}
\tablecaption{Molar Ratios of the Considered Elements in the Chondritic-like Material Relative to Si}
\label{tab:chonab}
\tablehead{\colhead{Element} & \colhead{C} & \colhead{O} & \colhead{Na} & \colhead{Mg} & \colhead{Si} & \colhead{Fe}}
\startdata
$N_i/N_{\ce{Si}}$ & 0.023 & 3.11 & 0.013 & 1.0 & 1.0 & 1.0 \\
\enddata
\end{deluxetable}
We consider two planets, one formed inside and one formed outside the water-ice line.
For both planets, the total planet mass is set to $M_\mathrm{pl}=4\,M_\oplus$ and the equilibrium temperature is $T_\mathrm{eq}=800\,\si{\K}$.
The sub-Neptune formed inside the water-ice line is composed of $96\%$ chondritic-like material and $4\%$ primordial gas. 
More specifically, the abundances of Si, Na, and O follow \citet{2001AllegreChemical}, and the molar ratios of Mg/Si and Fe/Si are parameterized and set to unity.
Table \ref{tab:chonab} lists the resulting elemental abundances of the chondritic-like material relative to Si.
The primordial gas is composed of $99.9\%$ \ce{H2} and $0.1\%$ \ce{CO2} by mole, corresponding to a solar C/O ratio \citep{2018Suarez-AndresCOMg}.
For simplicity, we use $f_{\ce{H2}}$ to refer to the mass fraction of primordial gas and $f_\mathrm{chon}$ to refer to the mass fraction of chondritic-like material throughout this work. 
In the case of a planet that formed outside the water-ice line, we assume a composition of $4\%$ primordial gas, $29\%$ \ce{H2O}, and $67\%$ chondritic-like material.

\begin{deluxetable*}{lcccccc}
\tablecaption{Overview over Model Inputs. $f_\mathrm{chon}$, $f_{\ce{H2}}$, $f_{\ce{H2O}}$ are the mass fraction of chondritic-like material, primordial, \ce{H2}-dominated gas, and water, respectively.}
\label{tab:inputparam}
\tablehead{\colhead{Planet} & \colhead{Mass [$M_\oplus$]} & \colhead{$f_\mathrm{chon}$ } &\colhead{$f_{\ce{H2}}$ } & \colhead{$f_{\ce{H2O}}$} & \colhead{$T_\mathrm{eq}$ [K]} & \colhead{$L_0$ [erg/s]}}
\startdata
Inside water-ice line  &  4 & 0.96 & 0.04 & 0.0 & 800 & $10^{22}$\\
Outside water-ice line & 4 & 0.67 & 0.04 & 0.29 & 800 &  $10^{22}$
\enddata
\end{deluxetable*}

For both planets, we compare the evolution including the atmosphere--interior coupling (\textit{chemically coupled case}) to the evolution without atmosphere--interior coupling (\textit{uncoupled case}). 
In all cases, we start the evolution at $t=0.1\,\si{\Gyr}$ with $L_0 = 10^{22}\,\si{\erg\per\s}$ and evolve the planet until $t=10\,\si{\Gyr}$ is reached.
The choice of the initial luminosity is based on the fit provided by \citet{mordasini_planetary_2020}. The input parameters are summarized in Table \ref{tab:inputparam}.

In the chemically coupled case, we use the global chemical equilibrium framework to calculate the atmospheric mass fraction and metallicity $Z$. 
The atmosphere mass corresponds to the mass of all species in the gas phase, while metallicity is defined as the mass fraction of all species other than \ce{H2} in the gas phase. 
The deep planetary interior is assumed to include all species in both the metallic and silicate phases, representing an extreme end-member case in which these phases remain well mixed rather than segregating into a core and mantle. In a differentiated planet, the metallic and silicate phases would instead correspond to the iron core and silicate mantle, respectively. 

Next to the elemental abundances, the temperature at the AMOI, which we take from the planetary structure model, is a key input parameter for the global chemical equilibrium framework. 
However, the $T_\mathrm{AMOI}$ is itself dependent on the mass and composition of the atmosphere. 
In order to calculate the atmosphere mass and composition self-consistently, we therefore use an iterative approach. 
We first calculate the planet structure, assuming all volatiles are confined to the atmosphere.
We use the resulting $T_\mathrm{AMOI}$ to calculate the atmosphere mass fraction and metallicity based on the global chemical equilibrium. 
Finally, we iterate between the global chemical equilibrium framework and the planet structure model using the updated atmospheric properties until we reach convergence. 

In order to evolve the planet, we calculate the energy at time $t + \Delta t$ based on the current total energy $E(t)$ and luminosity $L(t)$ using \cref{eq:dEdt}
\begin{equation}
    E(t + \Delta t) = E(t) - L_\mathrm{int}\Delta t + L_\mathrm{radio} \Delta t.
    \label{eq:dEdtapp}
\end{equation}
The time step $\Delta t$ is set by $\Delta t = c_{dt}|E_\mathrm{tot}(t)|/L(t)$ with $c_{dt} = 5\times 10^{-5}$ chosen to be small enough such that the approximation in \cref{eq:dEdtapp} holds. 
The luminosity at time $t+\Delta t$, $L_{t+\Delta t}$, is found by solving
\begin{equation}
    1 - \frac{E(t + \Delta t)}{E(L_{t + \Delta t})} = 0,
\end{equation}
for $L(t+\Delta t)$, which is the luminosity at the next time step. 
Here $E(L_{t + \Delta t})$ the total energy given by \cref{eq:Etot} based on the planetary structure calculated using $L(t+\Delta t)$.
In order to reduce the computational cost, we only update the atmosphere structure and $T_\mathrm{AMOI}$ when calculating $E(L_{t + \Delta t})$.
We thus assume that the contraction of the atmosphere is significantly stronger than the contraction of the interior, and fix the interior radius to its initial value. 
Consequently, the gravitational energy of the interior is constant in time and does not contribute to the luminosity of the planet.
Therefore, the assumption of a constant density for the interior will not affect our results.

Whenever the temperature at the AMOI has decreased by $\Delta T = 100\, \si{\K}$, we re-calculate the atmosphere mass fraction and metallicity with the global chemical equilibrium model using the iterative approach described above. 
The interior structure and radius are then updated using the new mass of the planet interior, which is given as $M_\mathrm{int} = M_\mathrm{pl} - M_\mathrm{atm}$. 

The planetary structure model assumes a differentiated interior with an Earth-like composition.
This is inconsistent with our assumption that the metallic phase continuously participates in atmosphere–interior exchange, which implies an undifferentiated interior. The effect of the differentiation state of a planet on its radius is negligible \citep{2025HuangDornlimits}.
By fixing the composition to be Earth-like, we further neglect to account for the sequestration of volatiles in the interior. 
The influence of this inconsistency is discussed in Section \ref{ssec:limitations}.
\section{Results}
\label{sec:results}
Our workflow allows us to compare sub-Neptunes formed inside and outside the water-ice line and assess how formation location, and thus initial volatile composition, affects properties such as their thermal evolution, contraction, atmospheric mass and composition.

\subsection{Evolution of the Uncoupled Case}
\begin{figure*}
    \centering
    \includegraphics[width=0.95\linewidth]{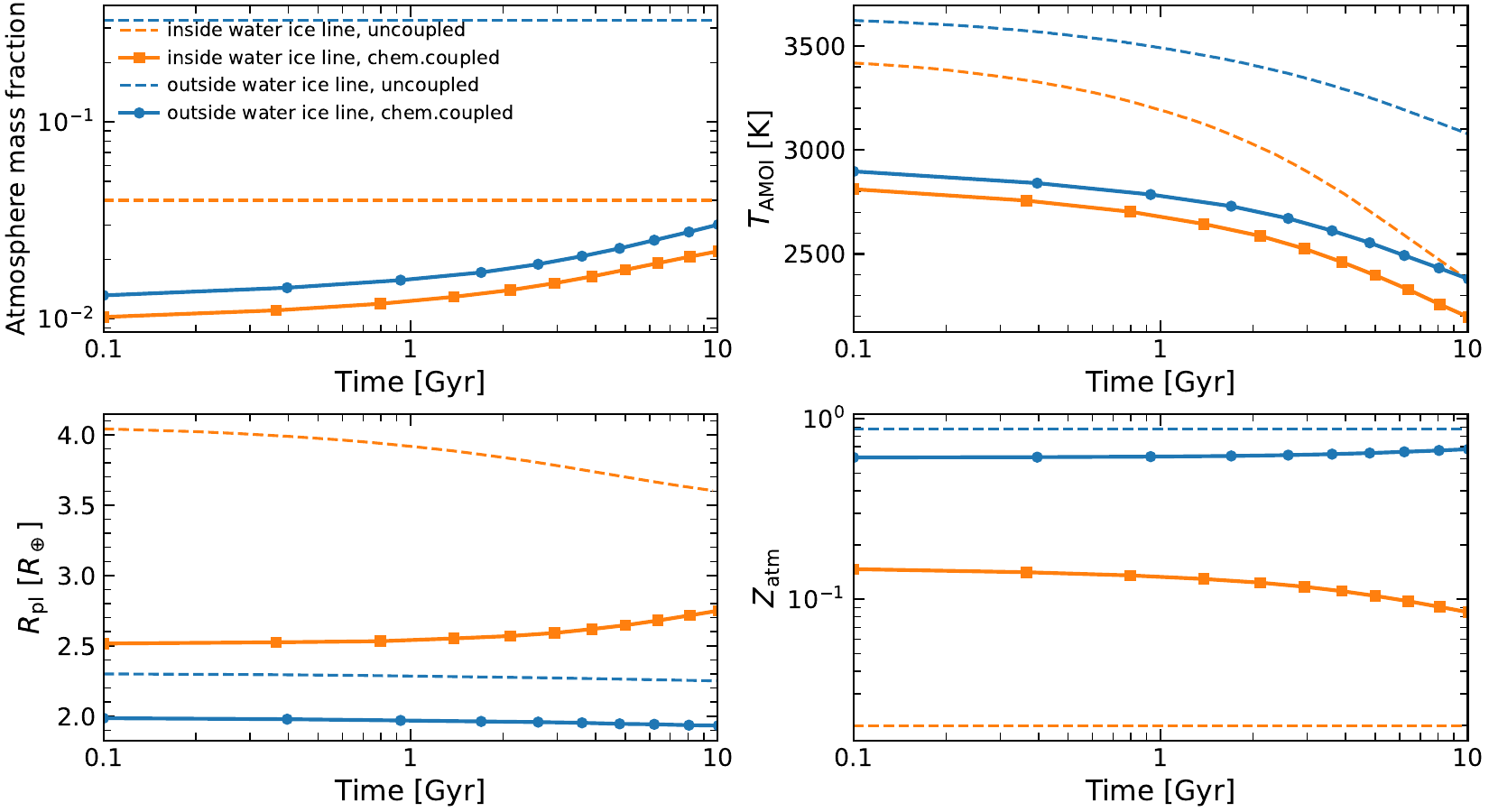}
    \caption{Comparison of the evolution of a $4\,M_\oplus$ planet with atmosphere--interior coupling (solid lines) and in the uncoupled case (dashed lines). The orange lines represent a sub-Neptune formed inside the water-ice line, while the blue lines correspond to a planet formed outside the water-ice line. Both planets have a total mass of $4\,M_\oplus$ and an equilibrium temperature of $T_\mathrm{eq}=800\,\si{\K}$. The top row shows the evolution of the atmosphere mass fraction and the temperature at the AMOI. The bottom row shows the evolution of the transit radius and the atmosphere metallicity. The symbols mark the snapshots at which the atmosphere mass and composition are recalculated in the chemically coupled case. For both planets, the outgassing of volatiles as the magma ocean cools leads to an increase in the atmosphere mass fraction. For the planet formed inside the water-ice line, the outgassing is strong enough to lead to a slight increase in radius with time and a decrease in the metallicity. The high accreted water abundance of the planet formed outside the water-ice line results in roughly constant radius and metallicity with time.}
    \label{fig:RplMatmZTAMOI}
\end{figure*}
We first discuss the evolution of the atmosphere mass fraction, the temperature at the AMOI, the transit radius, and the atmosphere metallicity for the two planets in the uncoupled case (dashed lines in Figure \ref{fig:RplMatmZTAMOI}). 

In this case, all volatiles are confined to the atmosphere of the planet, and the atmosphere mass fraction and metallicity remain constant throughout the evolution. 
For the planet formed dry (inside the water-ice line), this leads to an atmosphere mass fraction of $f_\mathrm{atm}=0.04$ with a metallicity of $Z_\mathrm{atm}=0.02$, while the atmosphere mass fraction of the planet formed outside the water-ice line is $f_\mathrm{atm}=0.33$ and the metallicity is $Z_\mathrm{atm}=0.88$.
In alignment with \citet{rogers_road_2025} and \citet{aguichine_evolution_2024}, we find that in the uncoupled case, the \ce{H2}-dominated atmosphere of the planet formed dry contracts as it cools, leading to a decrease in planet radius by 10\%, while the radius of the planet formed outside the water-ice line remains roughly constant in time (dashed lines in the bottom left plot in \Cref{fig:RplMatmZTAMOI}). 
The more massive atmosphere in the case of the planet formed outside the water-ice line increases the temperature at the AMOI and leads to slower cooling compared to the planet formed dry (dashed lines in the upper right plot in \Cref{fig:RplMatmZTAMOI}).

\subsection{Effect of Chemical Coupling on the Initial State}
The chemical interaction between atmosphere and interior in the coupled case leads to partitioning of volatiles into the interior as well as to a change in the atmosphere composition, which significantly alters the initial state and the consequent evolution of both planets.
We first concentrate on the change in the initial state compared to the uncoupled case and discuss the evolution in Section \ref{ssec:evolutioneffect}.

For the planet formed dry, these interactions reduce the initial atmosphere mass fraction to $f_\mathrm{atm}=0.010$. 
At the same time, the initial metallicity is increased to $Z_\mathrm{atm}=0.15$.
The reduction of the atmosphere mass fraction of the planet formed outside the water-ice line is even more pronounced. 
In this case, the initial atmosphere mass fraction is reduced to $f_\mathrm{atm}=0.013$ in the chemically coupled case, while the initial metallicity is $Z_\mathrm{atm}=0.61$. 

\begin{figure*}
    \centering
    \includegraphics[width=0.9\linewidth]{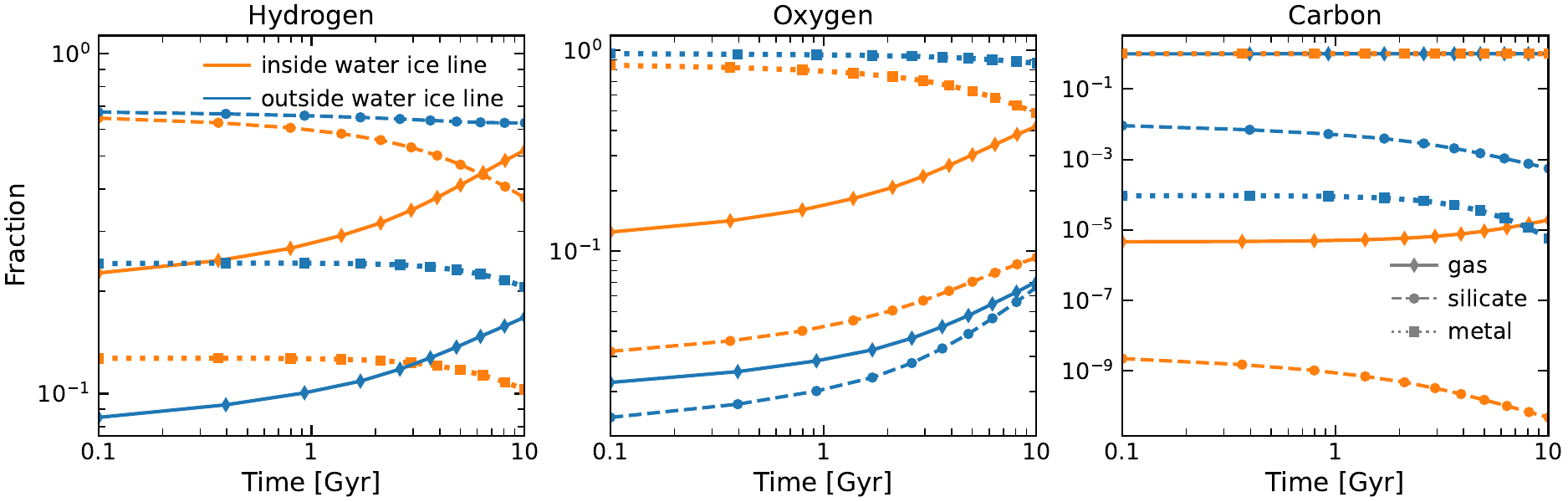}
    \caption{Partitioning of H (left), C (middle), and O (right) into metallic (dotted lines), silicate (dashed lines), and gaseous phases (solid lines) over time. The gas phase refers to the atmosphere of the planet, while the metallic and silicate phases refer to the deep planet interior. It is important to note that the fractions are normalized to the total abundance of each element and that only oxygen in non-silicate species is shown\footnote{Specifically, we only show oxygen in \ce{CO^{silicate}}, \ce{CO2^{silicate}}, and \ce{H2O^{silicate}}.}. The orange lines correspond to a planet formed inside the water-ice line, while the blue lines show a planet formed outside the water-ice line. In both cases, the total planet mass is $4\,M_\oplus$. 
    For the planet formed dry, carbon is predominantly in the metallic phase, while for the planet formed outside the water-ice line carbon is primarily in the gas phase. The fraction of C in the silicate phase for the planet formed dry is negligible.}  
    \label{fig:elemetndistr}
\end{figure*}

In order to understand the difference in the change of the initial atmosphere mass fraction between the two planet types, we examine the distribution of the key volatile elements H, C, and O between the gas, silicate, and metallic phases in \Cref{fig:elemetndistr}. In this figure, we only plot oxygen in non-silicate species\footnote{Specifically, we show oxygen in \ce{CO^{silicate}}, \ce{CO2^{silicate}}, and \ce{H2O^{silicate}}.}. 
For both planets, $\sim 65\%$ of hydrogen is initially dissolved in the silicate phase. 
However, for the planet formed dry, the remaining hydrogen is predominantly stored in the gas phase, whereas for the planet formed outside the water-ice line it is primarily stored in the metallic phase.
In contrast, the majority of the oxygen not bound in silicates is stored in the metallic phase in both planets, but for the planet formed outside the water-ice line $\sim 12\%$ of oxygen is located in the gas phase and $\sim 3\%$ in the silicate phase. 
For the planet formed dry only $2\%$ are in the gas phase and $1\%$ in the silicate phase.  
The high dissolution of hydrogen explains the reduction of the atmosphere mass fraction for the planet formed dry.
The significantly more pronounced reduction found in the planet formed outside the water-ice line is due to a higher total elemental abundance of oxygen and its strong sequestration into the metallic phase.

\begin{figure}
    \centering
    \includegraphics[width=0.95\linewidth]{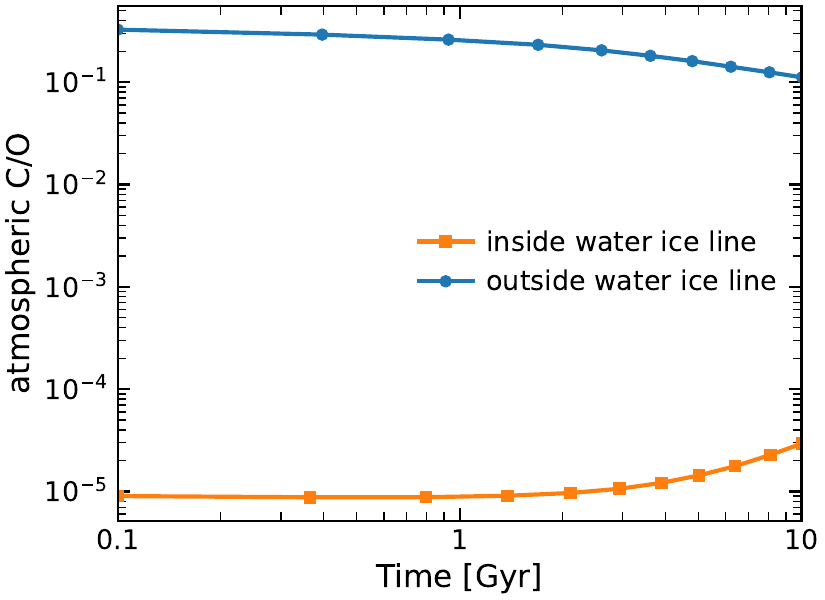}
    \caption{Evolution of the atmospheric C/O ratio for the planet formed dry (orange line) and the planet formed outside the water-ice line (blue line). The atmospheric C/O ratio differs by almost 4 orders of magnitude between the two formation locations at the end of the evolution.}
    \label{fig:COatmos}
\end{figure}

The partitioning behavior of carbon is very different for the planets from the two formation locations (\Cref{fig:elemetndistr}).
For the planet formed dry, C is stored predominantly in the metallic phase, while for the planet formed outside the water-ice line, C is stored predominantly in the gaseous phase. 
This leads to an initial atmospheric C/O ratio of $\sim 0.3$ for the planet formed outside the water-ice line, while the initial atmospheric C/O ratio of the planet formed dry is $\sim 10^{-5}$, see \Cref{fig:COatmos}.
Similar to \citet{werlen_atmospheric_2025}, we find that the atmosphere--interior interaction modifies the atmospheric C/O ratio compared to the C/O of the accreted primordial gas. 
Therefore, the atmospheric C/O can potentially be used to distinguish the two formation locations from one another.
Combined with the higher elemental abundance of O in the case of the planet formed outside the water-ice line, the different partitioning behavior of C for the two planets leads to a higher metallicity for the planet formed outside the water-ice line.

The lower atmosphere mass fraction in the chemically coupled case results in a significant reduction of $T_\mathrm{AMOI}$.
It further changes the initial radius of the planets. 
For the planet formed dry, the initial radius decreases from $R_\mathrm{pl}=4.04\,R_\oplus$ in the uncoupled case to $R_\mathrm{pl}=2.52\,R_\oplus$ in the chemically coupled case.
For the planet formed outside the water-ice line, the radius decreases from $R_\mathrm{pl}=2.30\,R_\oplus$ to $R_\mathrm{pl}=1.99\,R_\oplus$.

\subsection{Effect of Chemical Coupling on the Evolution}
\label{ssec:evolutioneffect}
Next, we examine how the distribution of the volatile elements change as the two planets cool, which is shown in \Cref{fig:elemetndistr}.

The cooling leads to a decrease in the temperature at the AMOI, as shown in the top right plot of \Cref{fig:RplMatmZTAMOI}. 
As a consequence, volatiles exsolve from the interior, causing the atmosphere mass fraction to increase with time for both planets. 
For the planet formed dry, the fraction of H in the metallic phase remains roughly constant, while the fraction of H in the silicate phase drops to 0.52.
The fraction of oxygen in the metallic phase decreases to 0.49 by the end of the evolution, leading to an increase of O in the gas and silicate phase.
The outgassing of H and O results in an increasing atmosphere mass fraction in time, with a final atmosphere mass fraction of 0.022.
This means that at the end of the evolution approximately half of the accreted hydrogen remains trapped in the interior for the planet formed dry.

For the planet formed outside the water-ice line, the final atmosphere mass fraction is $f_\mathrm{atm} = 0.030$, indicating that a much larger fraction ($\sim 90\%$) of the total volatile budget remains partitioned in the interior of the planet. 
This difference is again primarily due to the higher elemental abundance of oxygen and the different partitioning behavior of oxygen and hydrogen. 
Compared to the planet formed dry, there is less change in the volatile distribution over time.
While the fraction of hydrogen in the gas phase increases by a factor of 2 and the fraction of oxygen in the gas phase by a factor of 3, the majority of hydrogen and oxygen remain in the silicate and metallic phases, respectively.

This long-term exchange of volatile species between the metallic phase and the gas phase via the silicate phase is possible because we assume that the interiors of sub-Neptunes do not differentiate. For a differentiated planet, the participation of the metallic phase in the chemical interactions requires strong convection in the silicate phase \citep{2021Lichtenbergredox,schlichting2022}.

The evolution of the atmospheric metallicity and the C/O ratio differs significantly between the two formation locations. 
As the volatile budget of the planet formed dry is dominated by \ce{H} with only minor amounts of \ce{O} and \ce{C}, the metallicity decreases as hydrogen is outgassed over time, with a final metallicity of $Z_\mathrm{atm}=0.085$.
In contrast, the metallicity of the planet formed outside the water-ice line increases slightly with time, with a final metallicity of $Z_\mathrm{atm}=0.68$.
This is due to the higher total abundance of oxygen in this case. Further, C remains predominantly in the gas phase throughout the evolution of the planet formed outside the water-ice line.

For the planet formed outside the water-ice line, the increasing fraction of O in the gas phase as the planet evolves leads to a decrease in the atmospheric C/O ratio over time.
In contrast, the atmospheric C/O ratio of the planet formed inside the water-ice line increases slightly over time, due to a slight increase in C in the gas phase, see also \Cref{fig:elemetndistr}.
Although we do not include atmospheric escape, which could further affect the metallicity \citep{2026Valatsou}, the atmospheric C/O ratio remains a promising diagnostic of formation location in evolved planets, with planets formed dry exhibiting distinctly low C/O.
In addition, the atmospheric C/O ratio of the planet formed dry remains well below the C/O ratio of the accreted primordial gas throughout the evolution.

The chemically coupled case cools slower than the uncoupled case for the planet formed dry. 
This is caused by the higher optical thickness of the atmosphere due to the increase in the metallicity. 
The increase in atmosphere mass fraction due to the exsolution of hydrogen further raises the temperature at the AMOI.
The atmosphere of the planet formed outside the water-ice line is dominated by mass in \ce{H2O} both in the chemically coupled and uncoupled case. 
Consequently, the cooling behavior is similar in both cases, even though the temperature at the AMOI is decreased by $\sim 700\,\si{\K}$ in the chemically coupled case due to the lower atmosphere mass fraction.
Most importantly, the temperature at the AMOI remains well above $2000\,\si{\K}$ throughout the entire evolution of both planets. 
We therefore expect that the magma ocean will not solidify and the interior and atmosphere remain able to chemically interact at all times. 

Finally, we examine the effect of the change of the atmosphere mass fraction in time on the radius evolution of the two planets. 
During the planet's evolution, the increase in atmosphere mass due to the exsolution of volatiles can counteract the contraction of the planet radius caused by cooling. 
In combination with the high metallicity, this effect leads to an almost constant radius over the long-term evolution for the planet formed outside the water-ice line.
For the planet formed dry, the increase in atmosphere mass is strong enough to overcome the contraction, with a final radius of $R_\mathrm{pl}=2.75\,R_\oplus$.
Only when atmospheric loss is taken into account can it offset the exsolution effect, possibly leading to a net gradual decrease in planetary radius (see Section \ref{sec:Discussion}).
Our results highlight that it will not be possible to distinguish the two formation locations by the evolution of the radius or atmosphere mass fraction over time alone.

\subsection{Evolution of Gas Speciation}
\begin{figure*}
    \centering
    \includegraphics[width=0.9\linewidth]{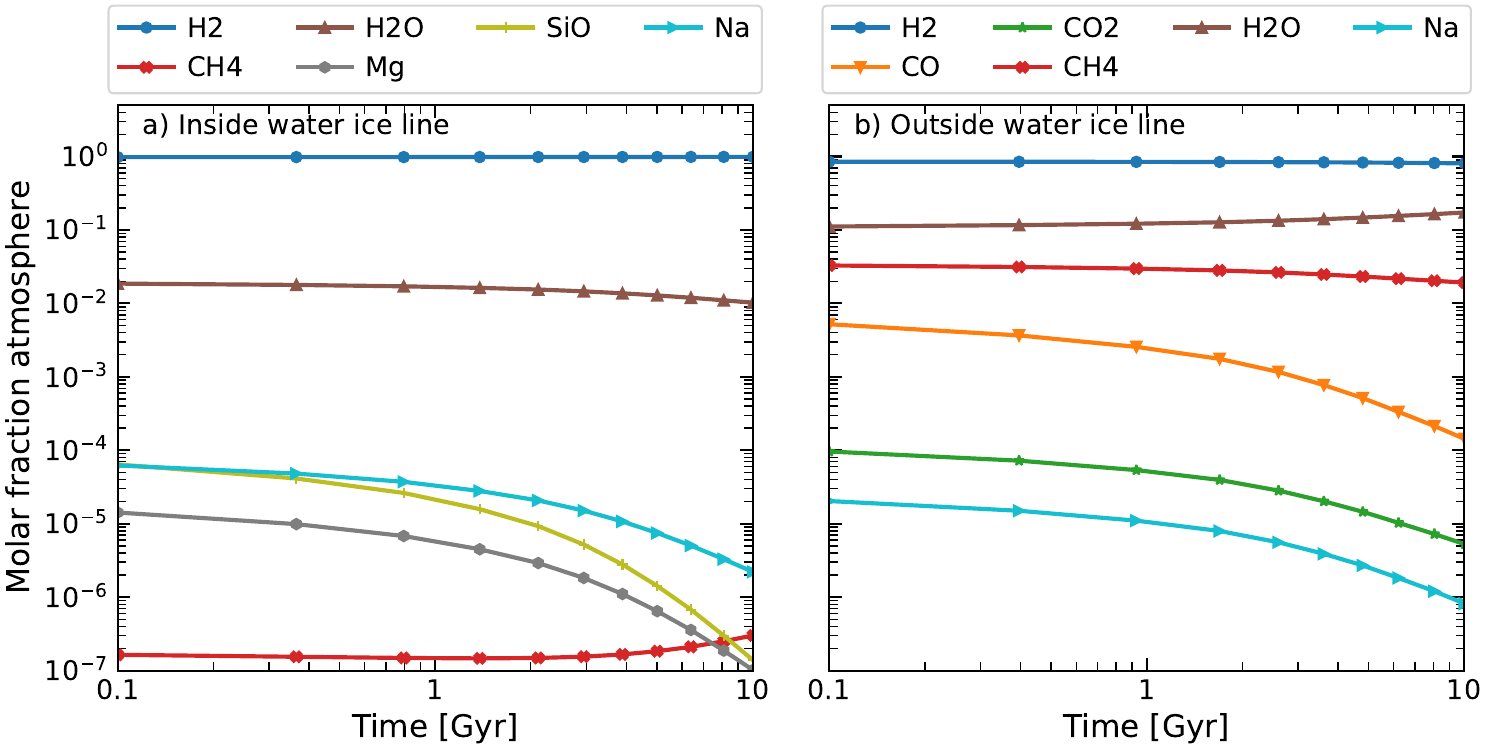}
    \caption{Evolution of the molar fractions of major atmospheric species over time. The left plot shows the atmosphere composition for a planet formed inside the water-ice line, and the right plot is for a planet formed outside the water-ice line. The planet mass is $4\,M_\oplus$ in both cases. For both planets, \ce{H2} and \ce{H2O} are the major atmospheric species. However, the mole fractions of \ce{CH4}, \ce{CO2}, and \ce{CO2} in the atmosphere of the planet formed outside the water-ice line are several orders of magnitude greater than the mole fractions in the planet formed inside the water-ice line.}
    \label{fig:atmospherecomp}
\end{figure*}
We now compare the molar fraction of the dominant gas species during the evolution of the two planets, shown in \Cref{fig:atmospherecomp} in more detail to determine the difference between the two formation locations. 
The key differences between the two formation locations are the molar fractions of carbon species in the atmosphere. 
For the planet formed dry, the most abundant carbon species is \ce{CH4}, with a molar fraction on the order of $10^{-7}$.
This is 5 orders of magnitude lower than the molar fraction of endogenically produced \ce{H2O}.
In contrast, the molar fraction of \ce{CH4} in the atmosphere of the planet formed outside the water-ice line is $\sim 10^{-2}$, which is less than an order of magnitude lower than the molar fraction of \ce{H2O}.
Furthermore, the fourth and fifth most abundant species in the atmosphere of the planet formed outside the water-ice line are \ce{CO} and \ce{CO2}, with initial abundances of $10^{-3}$ and $10^{-4}$ respectively.
However, the molar fractions of \ce{CO} and \ce{CO2} decrease by more than one order of magnitude as the planet evolves. 
The different molar fractions of the carbon species for the two formation locations are again explained by the different partitioning behavior of C for the two planets discussed above.
This demonstrates that the abundance of \ce{CH4} and other carbon species as well as their ratio to \ce{H2O} in the atmospheres of sub-Neptunes are potential diagnostics of their formation locations. 

\Cref{fig:atmospherecomp} further shows that, while the atmosphere of the planet formed dry is dominated by \ce{H2} and \ce{H2O}, especially at early times refractory species are stable in the gas phase, which is caused by the high AMOI temperatures (above $2000\,\si{\K}$). 

\section{Discussion}
\label{sec:Discussion}
Some formation models indicate that planets formed dry may not grow massive enough to capture enough primordial gas to become sub-Neptunes \citep{venturini_most_2020}.
Instead, the conditions required for the formation of sub-Neptunes may only be met beyond the water-ice line, where they will accrete both \ce{H2O} and \ce{H2} \citep{venturini_nature_2020}. Our results demonstrate that this connection can be probed observationally: atmospheric metallicity and molecular composition provide meaningful leverage for distinguishing formation locations, whereas radius evolution and atmospheric mass fraction offer comparatively little diagnostic power. 

More specifically, we argue that the abundance of \ce{CH4}, \ce{CO2}, and \ce{CO} in the atmosphere and the atmospheric C/O differ significantly between the two composition scenarios proposed for sub-Neptunes. 
In this section, we validate our results for a larger parameter space and model the vertical structure of the atmosphere in more detail.
We end with a comparison to the recent work by \citet{rogers_redefining_2025} and a discussion of the limitations of our model.
\subsection{Methane as Formation Tracer}
\begin{deluxetable}{lc}
\tablecaption{Parameter Range}
\label{tab:parameterrange}
\tablehead{\colhead{Parameter} & \colhead{Range}} 
\startdata
$M_\mathrm{pl} [M_\oplus]$ &  $1,2,4,6,8,10$ \\
$f_{\ce{H2O}}$ & $0.0, 0.1, 0.2, 0.5$\\
$f_{\ce{H2}}$ & $0.005, 0.01, 0.02, 0.03, 0.04, 0.05, 0.06, 0.07, 0.08, 0.09, 0.1$\\
$T_\mathrm{AMOI} [\si{\K}]$ & $2000, 3000$\\
$\Delta T [\si{\K}]$ & $50, 500, 1000$
\enddata
\end{deluxetable}
Our results suggest that a high methane abundance (molar fraction $>10^{-3}$) and a high atmospheric C/O ($\sim 1$) are indicative for high accreted water mass fractions and thus a formation outside the water-ice line.
In order to test whether this result is robust over a larger parameter space, we run a parameter study using the global chemical equilibrium model. 
Specifically, we vary the planet mass, the mass fraction of \ce{H2O} and \ce{H2}, $T_\mathrm{AMOI}$, and the temperature difference $\Delta T$ between the temperature at the AMOI and silicate - metal equilibrium temperature $T_\mathrm{SME}$, see Table \ref{tab:parameterrange}. 
All planets that have $f_{\ce{H2O}}\geq 0.1$ are assumed to have formed outside the water-ice line, while planets with $f_{\ce{H2O}}=0$ were formed dry. 
The ratio of Mg to Si and Fe to Si are fixed to 1 for all planets. 
\begin{figure*}
    \centering
 \includegraphics[width=0.95\linewidth]{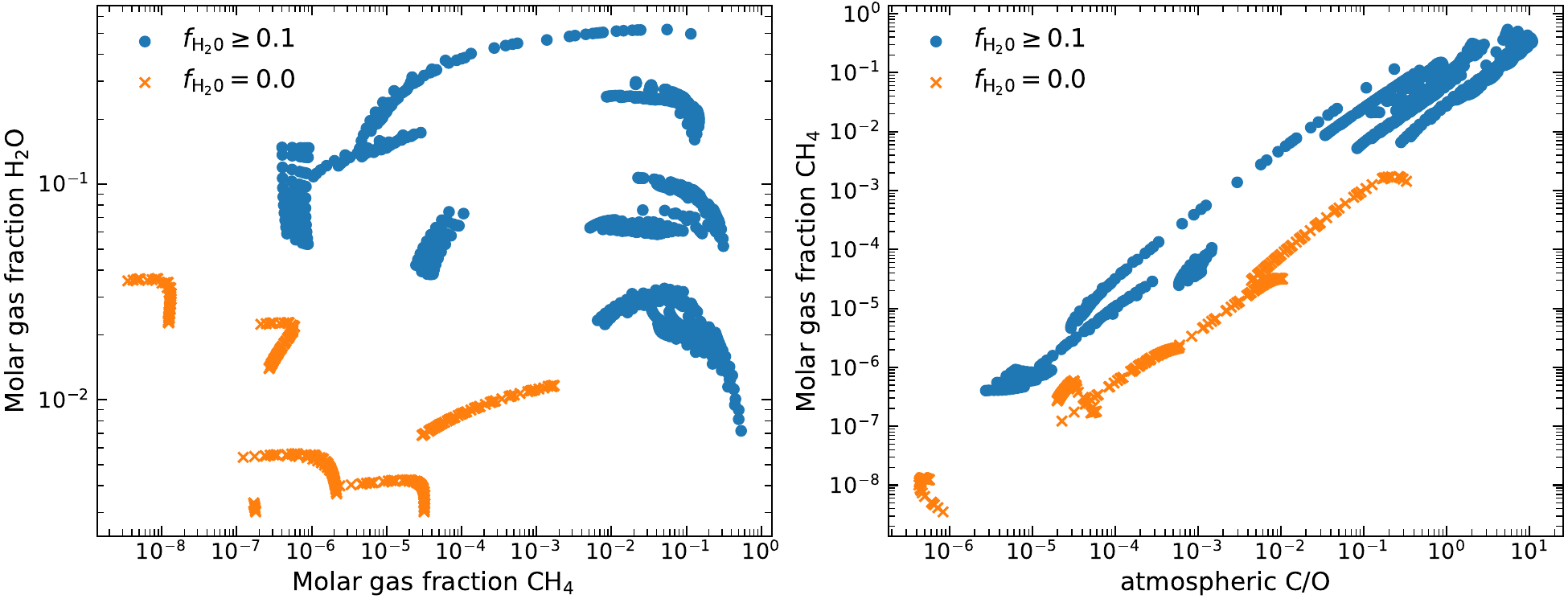}
    \caption{Molar gas fraction of \ce{H2O} versus \ce{CH4} (left plot) and molar gas fraction of \ce{CH4} versus atmospheric C/O ratio (right plot) for a synthetic population of planets (see Table \ref{tab:parameterrange}). The colors and symbols represent the accreted water mass content ($f_{\ce{H2O}}$) and thus indicate dry or water-rich formation. High \ce{CH4} abundances ($\geq10^{-2}$) and high atmospheric C/O ratios ($\geq$C/O$_{solar}$) are only found in atmospheres of planets formed water-rich ($f_{\ce{H2O}}\geq 0.1$).
    }
    \label{fig:parameterstudy}
\end{figure*}

The left plot in \Cref{fig:parameterstudy} shows the molar gas fractions of \ce{H2O} and \ce{CH4} as a function of accreted water mass content (left plot), while the right plot in \Cref{fig:parameterstudy} shows the molar gas fraction of \ce{CH4} and the atmospheric C/O ratio.
Indeed, we find that a molar gas fraction of \ce{CH4} $\gtrsim 5\times 10^{-3}$ is only possible for planets that accreted a water mass fraction $\geq 10\%$ (blue dots in \Cref{fig:parameterstudy}).
In contrast, the molar gas fraction of \ce{CH4} for the majority of planets that did not accrete any water is below $10^{-3}$ (orange points in \Cref{fig:parameterstudy}). 
Similarly, the molar fraction of \ce{H2O} in the atmosphere of these planets is $\lesssim 5 \times 10^{-2}$, while planets that have accreted water can have a molar fraction of \ce{H2O} as high as 0.5.

A low molar gas fraction of \ce{CH4} cannot be used to rule out a formation outside the water-ice line.
Across a wide range of thermal histories and water accretion fractions, planets formed beyond the ice line can exhibit a large range of methane abundances in the atmosphere, ranging from $10^{-7}$ to $10^{-1}$ (Figure \ref{fig:parameterstudy}).
Although planets formed inside and outside the water-ice line can have a wide range of possible atmospheric C/O ratios, solar to supersolar C/O ratios ($\gtrsim0.55$) are mainly reached in planets that have accreted water, see right plot in \Cref{fig:parameterstudy}. 

The parameter study confirms that, to first approximation, molar gas fractions larger than $10^{-1}$ for \ce{H2O} and larger than $10^{-2}$ for \ce{CH4} respectively serve as strong indicators for water-rich accretion and thus formation beyond the water-ice line. 

\subsection{Vertical Structure}

\begin{figure}
    \centering
    \includegraphics[width=0.9\linewidth]{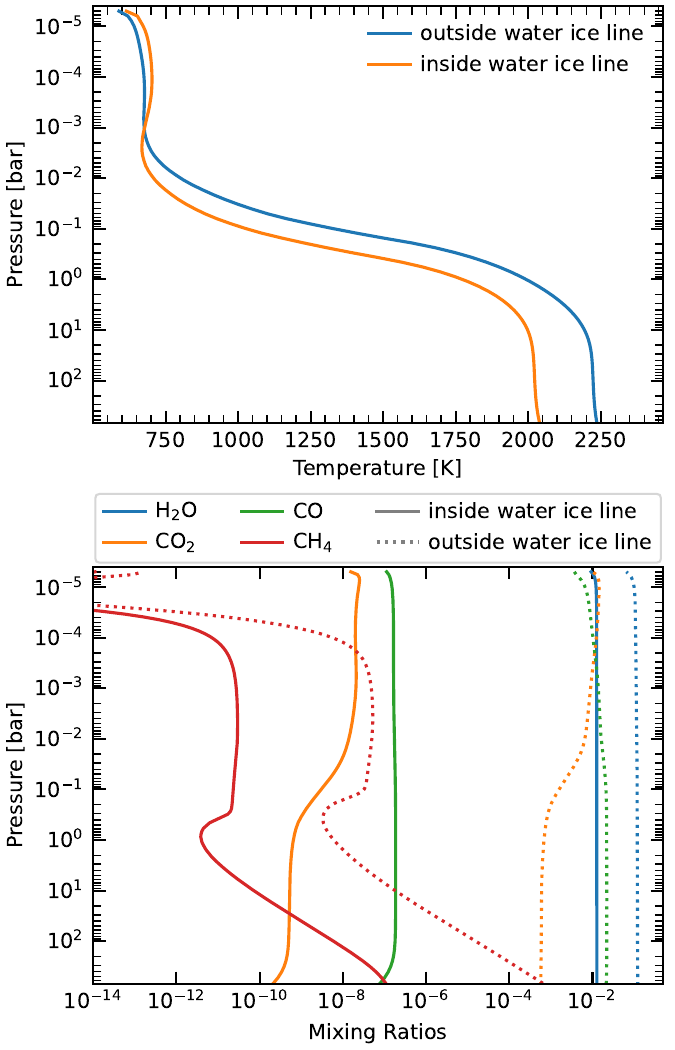}
    \caption{Structure of the upper atmosphere at $5\,\si{\Gyr}$. Top: pressure–temperature profiles (blue: formed outside the water-ice line; orange: formed inside). Bottom: vertical mixing ratios of dominant C- and O-bearing species, assuming $K_\mathrm{zz}=10^8,\si{\cm\squared\per\s}$. Solid lines show the planet formed dry; dotted lines show the planet formed water-rich. The C/O ratio remains constant in the upper atmosphere.}
    \label{fig:verticalmixing}
\end{figure}
We have demonstrated that the abundance of carbon species in the atmosphere depends on the accreted water abundance of the planet.
However, this result focuses on the composition of the lower atmosphere near the AMOI.
Spectroscopic observations only probe the upper atmosphere, which is expected to have a different composition due to variations in the thermochemical equilibrium state caused by the decrease in temperature and pressure with altitude, as well as by vertical mixing and photochemistry.
Therefore, to distinguish different bulk compositions by observations, it is necessary to connect the chemical composition calculated at the AMOI to the composition of the upper atmosphere.
To achieve this goal, we couple a suite of open-source models: \texttt{FastChem} \citep{stock_fastchem_2018}, \texttt{HELIOS} \citep{malik_helios_2017, malik_self-luminous_2019}, \texttt{VULCAN} \citep{tsai_vulcan_2017}, and \texttt{HELIOS-K} \citep{grimm_helios-k_2021}. Together, these codes capture gas-phase chemical equilibrium, radiative–convective transport, photochemistry, and wavelength-dependent opacities across a wide pressure range.

\texttt{FastChem} computes local chemical equilibrium in the gas phase. \texttt{HELIOS} solves the 1D radiative–convective structure, using opacities calculated by \texttt{HELIOS-K}. \texttt{VULCAN} integrates chemical kinetics, including thermochemistry, photochemistry, vertical mixing, and condensation. Starting from the equilibrium composition obtained at the AMOI, we use \texttt{FastChem} to compute equilibrium mixing ratios on a pressure–temperature grid. These are passed to \texttt{HELIOS-K} to calculate opacities, which then serve as inputs for \texttt{HELIOS} to obtain the atmospheric P–T profile. The P–T profile, along with the AMOI composition as a lower boundary condition, is subsequently supplied to \texttt{VULCAN} to compute the steady-state chemical structure. We then iterate between \texttt{VULCAN}, \texttt{HELIOS-K}, and \texttt{HELIOS} until both mixing ratios and the P–T profile converge.  
The \texttt{HELIOS} grid extends from $10^{4}\,$bars to $10^{-5}\,$bars, while \texttt{VULCAN} integrates kinetics from $10^{3}\,$bars upward, assuming perfect mixing in the deep convective region.
The strength of the vertical mixing in sub-Neptunes remains unconstrained \citep[e.g.][]{2015Charnay3D,2021TsaiInferring}.
We set $K_\mathrm{zz}=10^8\,\si{\cm\squared\per\s}$ in this study.
However, \citet{2025Nixonmagma} showed that the volume mixing ratios of \ce{H2O}, \ce{CH4}, and \ce{CO} are largely unaffected by the strength of vertical mixing for $K_\mathrm{zz}$ in the range of $10^{5}\,\si{\cm\squared\per\s}$ to $10^{9}\,\si{\cm\squared\per\s}$. 

\Cref{fig:verticalmixing} shows that the dominant C- and O-bearing species exhibit similar vertical behavior in both planets at $5\,\si{\Gyr}$. The mixing ratio of \ce{H2O} remains nearly constant throughout the atmospheric profile. As we go from the deeper atmosphere upwards, \ce{CO} replaces \ce{CH4} as the dominant carbon species. 
The mixing ratio of \ce{CO2} increases with altitude for both planets. 
For the planet formed dry, \ce{CO2} becomes the second most abundant C-bearing species for $P<10$\,bars.
For the planet formed outside the water-ice line, \ce{CO2} even becomes the most abundance C-bearing species for $P<10^{-4}\,$bars.
The mixing ratio of \ce{CO2} would be likely enhanced further for weaker mixing ($K_\mathrm{zz}<10^8\,\si{\cm^2,s^{-1}}$).
Crucially, the upper-atmosphere composition preserves the deep-atmosphere C/O ratio. 
The two formation locations remain distinguishable by the atmospheric composition: outside the water-ice line, \ce{H2O} and \ce{CO} have comparable mixing ratios, whereas inside the water-ice line \ce{CO} is several orders of magnitude lower than \ce{H2O}. 
Thus, the compositional differences identified in \Cref{fig:COatmos} and \Cref{fig:atmospherecomp} persist into the observable upper atmosphere, even though the dominant carbon carriers shift with altitude.

\subsection{Comparison to J. G. Rogers et al. (2025)}
Recently, \citet{rogers_redefining_2025} studied the evolution of the structure of sub-Neptunes taking into account the miscibility of \ce{H2} and \ce{MgSiO3} based on ab initio - molecular dynamics calculations by \citet{gilmore_core-envelope_2026}. 
They define the surface of the planet as the binodal surface, which marks the phase transition between miscible and immiscible. For young sub-Neptunes, they find that most hydrogen resides in the miscible interior, reducing both planet radius and envelope mass fraction relative to classical models. As the planet cools, the binodal temperature falls and hydrogen progressively exsolves; after $\sim$Gyr timescales, only a small residual fraction remains in the interior.

In contrast, our work examines how atmosphere–interior chemical exchange influences sub-Neptune evolution using a global chemical-equilibrium framework. Like \citet{rogers_redefining_2025}, we find that most volatiles are initially sequestered in the interior. 
However, our results show that approximately half of the accreted hydrogen remains sequestered in the interior of the planet formed dry at the end of the evolution.
We further find that the majority of oxygen and a small fraction of hydrogen remain in the metallic phase.
This highlights the metallic phase as a significant long-term volatile reservoir, which is absent from \citet{rogers_redefining_2025}. 
Nevertheless, by neglecting \ce{H2}–\ce{MgSiO3} miscibility, our framework likely underestimates interior hydrogen storage, particularly for young planets.

We emphasize that, unlike \citet{rogers_redefining_2025}, our framework explicitly tracks chemical exchange between the atmosphere and interior. As a result, the atmospheric composition is not fixed but can evolve over time as volatiles redistribute between phases.

\subsection{Model Limitations}
\label{ssec:limitations}
Our model relies on several assumptions discussed in the following.

The initial luminosity is one of the key input parameters for the evolution of sub-Neptunes. 
In this work, we use $L_0 = 10^{22}\,\si{\erg\per\s}$, which is lower than the initial luminosities used in other evolution studies \citep[e.g., ][]{2025_Tang_sub-Neptune,rogers_redefining_2025}.
Increasing the initial luminosity to values $L_0>10^{22}\,\si{\erg\per\s}$ results in $T_\mathrm{AMOI}>4000\,\si{\K}$. 
However, calculating the global chemical equilibrium at such high temperatures is beyond its capabilities due to limits in the thermodynamic database.

Our thermal--chemical evolution model assumes that the timescale needed for the planet to reach chemical equilibrium is shorter than the cooling timescale of the planet. 
Future work is needed to test if this assumption is valid.

Furthermore, our model does not account for the potential emergence of layered convection within the interior and atmosphere of the planet. 
While vigorous convective mixing within silicate melt is a standard assumption in the literature \citep[e.g.,][]{2021Lichtenbergredox,2024Nicholss_magma,2024_Seo_subNeptune}, different regimes, such as sluggish or double-layered convection, have also been proposed to occur in rocky exoplanets \citep[e.g.,][]{2012Stamenkovicinfluence,2020Spaargaren}.
Similarly, mean molecular weight gradients have been shown to inhibit convection in the deep atmosphere of sub-Neptunes \citep[e.g.,][]{2017Leconte,2022Misener,2022Markham,2024Leconte3d}.
The occurrence of layered convection could prevent the planet from reaching the state of global chemical equilibrium predicted by our model. 
However, future work (including 2D or 3D models) is required to determine under which circumstances long-timescale mixing is enabled or suppressed in both the interiors and atmospheres of sub-Neptunes.

In this study, we consider the case of an undifferentiated planet for which the metallic phase remains chemically coupled to the silicate and gaseous phase throughout the planet's evolution. 
In a differentiated planet, however, the metallic phase can be chemically isolated and hence, only silicate and gaseous phases may achieve chemical equilibrium.
Nevertheless, we anticipate that during the core formation process a significant fraction of volatiles will be partitioned into the metallic phase, consistent with the initial distribution predicted by our global chemical equilibrium model.
Consequently, the significant difference in the abundance of gaseous carbon species between the two formation locations is expected to remain a robust signature, even if the metallic core is excluded from the long-term chemical exchange. 
That said, such planets likely exhibit a more pronounced reduction in the atmospheric metallicity over time, independent of their formation location.
Oxygen partitioned in the metallic phase will be mainly trapped in the core and not transition back to the gas phase as the planet cools, while hydrogen will still outgas from the silicate phase. 
Especially for the planet formed outside the water-ice line, this could further reduce the final atmospheric mass fraction.

Our structure model assumes a differentiated interior with an Earth-like silicate mantle and iron core. This formally conflicts with our assumption that the metallic phase continuously participates in atmosphere–interior exchange, which would imply an undifferentiated interior. However, for our purposes the only relevant interior property is its radius, and the difference between differentiated and undifferentiated configurations is expected to be $\sim1\%$ \citep[][and Luo et al. (submitted)]{2025HuangDornlimits}, and thus too small to affect our results.
A more critical limitation is that the interior model is not yet capable of accurately accounting for the large amount of volatiles sequestered into the interior of the planet.
By assuming an Earth-like composition, we likely overestimate the radius of the interior, however only small changes are expected \citep{2025young_differentiation}.
More specifically, for the planet formed outside the water-ice line, the radii shown in \Cref{fig:RplMatmZTAMOI} may be somewhat overestimated due to the large amount of oxygen partitioned into the metallic phase. 
As the envelopes dominate sub-Neptune radii, we anticipate that the qualitative behavior of the radius evolution, as well as the evolution of the atmosphere mass and composition, remain robust. 

This study does not consider atmospheric escape, which plays an important role in shaping the evolution of sub-Neptunes \citep[e.g.,][]{owen_kepler_2013,jin_planetary_2014,owen_evaporation_2017,misener_cool_2021,2025_Tang_sub-Neptune}. 
The loss of atmosphere due to escape leads to a significant decrease of planet radius with time. 
This effect is likely stronger for the planet formed dry due to the lower atmospheric metallicity.

\section{Conclusions}
\label{sec:conclusions}
In this work, we present the first thermal--chemical planet evolution models that take into account the chemical interaction between the atmosphere and interior, focusing on sub-Neptunes. 
atmosphere--interior coupling significantly affects the evolution of sub-Neptunes, specifically the evolution of their radii, atmosphere mass fractions, and atmosphere compositions.  We highlight the implications for a water-rich sub-Neptune formed outside the water-ice line and a sub-Neptune formed dry (inside the water-ice line). 
We identify observational signatures that allow us to distinguish these two types of sub-Neptunes. 

Chemical coupling enables the partitioning of volatiles deep within planetary interiors. As a result, atmospheric masses can be significantly reduced, especially during early evolution. As planets cool, volatiles (primarily hydrogen) may subsequently exsolve from the interior and replenish the atmosphere.
For both formation locations, the increase in atmosphere mass fraction can counteract the thermal contraction of the planet, leading to an almost constant radius in time. 
In contrast to the hypothesis proposed by \citet{aguichine_evolution_2024} and \citet{rogers_road_2025}, we conclude that the radius evolution of sub-Neptunes alone is insufficient to distinguish between different composition types.

Instead, we find that the two formation scenarios can be distinguished by their atmospheric compositions.
While the endogenic production of water raises the atmospheric metallicity of the planet formed dry, it remains well below that of the planet formed outside the water-ice line.
By the end of its evolution, the metallicity of the planet formed dry is a factor of 8 lower than the metallicity of the planet formed outside of the water-ice line.

The most significant difference between the two formation scenarios is the abundance of carbon species in the atmosphere and the resulting atmospheric C/O ratio.
For a sub-Neptune formed dry, the atmospheric \ce{CH4} abundance, its dominant C-bearing gas species, is almost 5 orders of magnitude lower than its \ce{H2O} abundance. 
In contrast, the abundances of \ce{CH4} and \ce{H2O} in a planet formed outside the water-ice line differ by less than one order of magnitude.
Using a broad parameter study, we confirm that a high atmospheric abundance of \ce{CH4} ($\>10^{-2}$) or \ce{H2O} ($> 5\times10^{-2}$), or super-solar C/O ratios are a unique feature of planets formed beyond the water-ice line across different planet masses and compositions.
At the same time, sub-Neptunes can exhibit a broad range of \ce{CH4}abundances of \ce{CH4} and C/O ratios, regardless of their formation location.

The use of the atmospheric C/O ratio as a tracer of a planet's formation location was first proposed by \citet{2011Obergeffect} in the context of gas giants. 
In this scenario, the C/O in the atmosphere of a gas giant is directly inherited from its formation location relative to the ice lines of different volatile species in the protoplanetary disk. 
Later studies, incorporating more physically and chemically consistent models of planet formation, have shown that the atmospheric C/O ratio of gas giants is further influenced by various processes, such as planetesimal accretion \citep[e.g,][]{2016Mordasini,2020Shibata} or the enrichment of disk gas by the sublimation of inward-drifting pebbles \citep[e.g.,][]{2021Schneider}.
However, it is important to note that the atmospheric C/O ratios in sub-Neptunes are not directly inherited from formation, but significantly altered by chemical interaction between the planetary interior and the atmosphere \citep{werlen_atmospheric_2025}. 
Since these interactions depend on the bulk volatile composition of a planet, the atmospheric C/O ratios may serve as a potential signature of their formation location, provided they are interpreted using chemically coupled models.

The James Webb Space Telescope has provided us with the first well-characterized atmospheres of sub-Neptunes \citep[e.g., ][]{benneke_jwst_2024,piaulet-ghorayeb_jwstniriss_2024,2025DavenportTOI,2025SchmidtK218b}. 
This allowed, for the first time, the measurement of the atmospheric C/O ratio of a sub-Neptune, namely TOI-270\,d \citep{benneke_jwst_2024,2025FelixCompeting}. 
Their measured value for C/O is of the same order of magnitude as our prediction for a formation outside the water-ice line in line with current formation models \citep{venturini_nature_2020}.

The number of well-characterized sub-Neptune atmospheres spanning a wide range of equilibrium temperatures is expected to increase significantly in the coming years with the launch of the Ariel mission \citep{2022TinettiAriel}. 
Our results show that, although atmospheric composition does not directly reflect the total accreted volatiles of sub-Neptunes, it remains a powerful diagnostic of accreted volatile composition and thus their formation location. 

Finally, consistent with previous studies \citep{dorn_hidden_2021,luo_majority_2024,werlen_sub-neptunes_2025}, we find that the majority of the bulk volatile budget of a sub-Neptunes is stored in the interior of the planet and not in the atmosphere. 
This emphasizes the need for chemically and compositionally coupled models for the interpretation of mass-radius data in order to not underestimate the total volatile budget of a planet. This is relevant for all transit missions (e.g., Kepler, TESS, CHEOPS, PLATO) and mass follow-up characterizations.

\begin{acknowledgments}
We thank the referee for the constructive comments.
We thank Allona Vazan for providing the evolution curves used for comparison in the appendix.
CD gratefully acknowledges support from the Swiss National Science Foundation under grant TMSGI2\_211313. This work has been carried out within the framework of the NCCR PlanetS supported by the Swiss National Science Foundation under grant 51NF40\_205606.
We acknowledge the use of large language models (LLMs) to improve
the grammar, clarity, and readability of the manuscript. After using this service, we reviewed and edited the content as needed and take full responsibility for the content of the publication. 
\end{acknowledgments}

% \facilities{HST(STIS), Swift(XRT and UVOT)}
\software{numpy \citep{numpy}, matplotlib \citep{matplotlib}, pandas \citep{reback2020pandas}, h5py \citep{collette_h5pyh5py_2022}
          }
\appendix
\restartappendixnumbering
\section{Evolution Benchmarking}
\label{app:benchmark}

We benchmark our evolutionary framework by comparing the time evolution of planetary luminosity to that predicted by the model of \citet{vazan_convection_2015,vazan_contribution_2018} (hereafter the VZ model).
In addition to the structure equations \cref{eq:presstruceq}- \cref{eq:structureeq}, this model solves the equation for energy balance 
\begin{equation}
    \frac{\partial u}{\partial t} + p \frac{\partial}{\partial t} \frac{1}{\rho} = q - \frac{\partial L}{\partial m}, 
\end{equation}
where $u$, $p$, $\rho$ are the specific energy, the pressure, and the density respectively. The contribution of additional energy sources such as radioactive decay is given by $q$, whereas $m$ and $L$ are the mass and luminosity. 

We test three different planet masses, $4$, $5$, and $6\,M_\oplus$. For the $5\,M_\oplus$ planet we test two different atmosphere mass fractions, $f_\mathrm{atm} = 0.026$ and $f_\mathrm{atm}=0.05$, while for the other two planets we only test $f_\mathrm{atm} = 0.026$. 
For all planets, the equilibrium temperature is set to $T_\mathrm{eq}\approx 620\,\si{\K}$.

\begin{figure}
    \centering
    \includegraphics[width=0.95\linewidth]{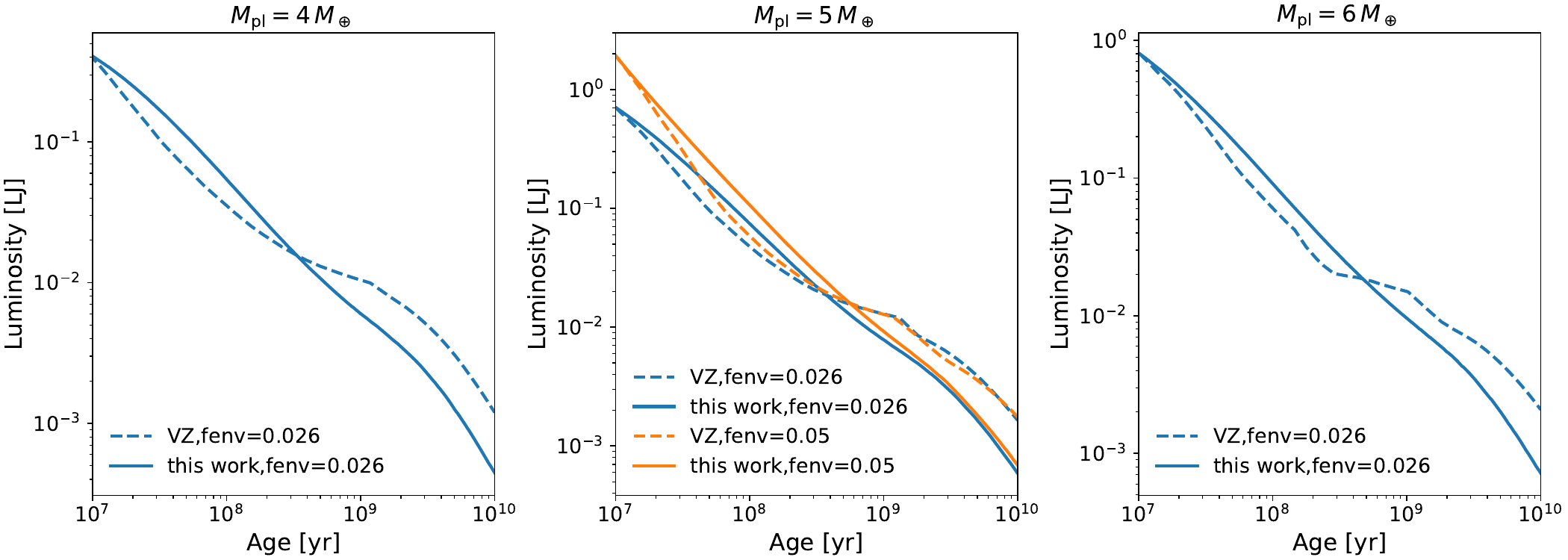}
    \caption{Comparison of the luminosity as a function of time calculated with our evolution framework (solid lines) and to luminosity calculated by the VZ evolution model {dashed lines} for three different planet masses and an atmosphere mass fraction of $f_\mathrm{atm}=0.026$ (blue lines). For the planet with $M_\mathrm{pl} = 5\,M_\oplus$, we further consider an atmosphere mass fraction of $f_\mathrm{atm} = 0.05$ (orange lines). The equilibrium temperature is set to $T_\mathrm{eq}\approx 620\,\si{\K}$ for all planets.}
    \label{fig:benchmarkLum}
\end{figure}

The resulting cooling curves are shown in \Cref{fig:benchmarkLum}.
The qualitative behavior of the curves calculated with our evolution framework is similar to the ones of the VZ model, even though the methodology is quite different. 
However, the evolution framework presented in this work exhibits lower luminosities and stronger cooling for $t\gtrsim3\times 10^{8}\,$yr. 
This difference is likely caused by the different treatment of the planet interior and the radioactive heating in the two models. 
Our model assumes an Earth-like interior, where only the mantle is contributing to the radioactive luminosity, as described in \citep{mordasini_characterization_II_2012}. 
In the VZ model, the interior is described using the thermodynamic properties of \ce{SiO2} and the values for the radioactive heating are taken from \citet{nettelmann_thermal_2011}.
This results in a larger contribution to the total energy of the planet by radioactive decay in the VZ model, which slows down the cooling of the planet.

%% the bibliography. The Main.bib file was populated from ADS. To
%% pdflatex Main.tex
%% bibtext Main
%% pdflatex Main.tex
%% pdflatex Main.tex

\bibliography{references}{}
\bibliographystyle{aasjournalv7}



\end{document}